\documentclass[english,aps,prl,twocolumn,showpacs,superscriptaddress,,nofootinbibfloatfix]{revtex4-1}
\usepackage[T1]{fontenc}
\usepackage[latin1]{inputenc}
\usepackage{graphicx}
\usepackage{bm}
\usepackage{amssymb}
\usepackage{color}
\usepackage[usenames,dvipsnames]{xcolor}
\usepackage{amsmath}
\usepackage{amstext}
\usepackage{latexsym}
\usepackage[colorlinks=true,citecolor=Cerulean,linkcolor=RubineRed,urlcolor=Cerulean]{hyperref}

\usepackage{mathptmx} 
\usepackage{verbatim}

\usepackage{amsfonts}
\usepackage{epsfig}
\usepackage{dsfont}
\usepackage{mathrsfs}
\usepackage{arydshln,leftidx,mathtools}
\usepackage{enumitem}

\usepackage{color}

\begin{document}

\title{An Interaction-Driven Many-Particle Quantum Heat Engine: Universal Behavior}

\author{Yang-Yang Chen}

\affiliation{Shenzhen Institute for Quantum Science and Engineering, and Department of Physics, Southern University of Science and Technology, Shenzhen 518055, China}
\affiliation{State Key Laboratory of Magnetic Resonance and Atomic and Molecular Physics,
Wuhan Institute of Physics and Mathematics, Chinese Academy of Sciences, Wuhan 430071, China}

\author{Gentaro Watanabe}
\affiliation{Department of Physics and Zhejiang Institute of Modern Physics,
Zhejiang University, Hangzhou, Zhejiang 310027, China}

\author{Yi-Cong Yu}
\affiliation{State Key Laboratory of Magnetic Resonance and Atomic and Molecular Physics,
Wuhan Institute of Physics and Mathematics, Chinese Academy of Sciences, Wuhan 430071, China}

\author{Xi-Wen Guan}
\email{xwe105@wipm.ac.cn}
\affiliation{State Key Laboratory of Magnetic Resonance and Atomic and Molecular Physics,
Wuhan Institute of Physics and Mathematics, Chinese Academy of Sciences, Wuhan 430071, China}
\affiliation{Department of Theoretical Physics, Research School of Physics and Engineering,
Australian National University, Canberra ACT 0200, Australia}

\author{Adolfo del Campo}
\email{adolfo.delcampo@umb.edu}
\affiliation{Department of Physics, University of Massachusetts, Boston, MA 02125, USA}
\affiliation{Theoretical Division, Los Alamos National Laboratory, Los Alamos, NM 87545, USA}

\def\q{{\bf q}}

\def\G{\Gamma}
\def\L{\Lambda}
\def\la{\lambda}
\def\g{\gamma}
\def\al{\alpha}
\def\s{\sigma}
\def\e{\epsilon}
\def\k{\kappa}
\def\ve{\varepsilon}
\def\l{\left}
\def\r{\right}
\def\te{\mbox{e}}
\def\d{{\rm d}}
\def\t{{\rm t}}
\def\K{{\rm K}}
\def\N{{\rm N}}
\def\H{{\rm H}}
\def\la{\langle}
\def\ra{\rangle}
\def\om{\omega}
\def\Om{\Omega}
\def\vep{\varepsilon}
\def\wh{\widehat}
\def\tr{{\rm Tr}}
\def\da{\dagger}
\def\iz{\left}
\def\zi{\right}
\newcommand{\beq}{\begin{equation}}
\newcommand{\eeq}{\end{equation}}
\newcommand{\beqa}{\begin{eqnarray}}
\newcommand{\eeqa}{\end{eqnarray}}
\newcommand{\intf}{\int_{-\infty}^\infty}
\newcommand{\into}{\int_0^\infty}

\begin{abstract}

A quantum heat engine (QHE) based on the interaction driving of a many-particle working medium is introduced.
The cycle alternates isochoric heating and cooling strokes with both interaction-driven processes that are simultaneously isochoric and isentropic.
When the working substance is confined in a tight waveguide, the  efficiency of the cycle becomes universal at low temperatures and governed by the ratio of velocities of a Luttinger liquid. We demonstrate the performance of the engine with an interacting Bose gas as a working medium and show that the average work per particle is maximum at criticality. We further discuss a work outcoupling mechanism based on the dependence of the interaction strength on the external spin degrees of freedom.
\end{abstract}

\maketitle

Universality plays a crucial role in thermodynamics, as emphasized by the description of heat engines, that transform heat and other resources into work.
In paradigmatic cycles (Carnot, Otto, etc.)  the role of the working substance is secondary.
The emphasis on universality thus seems to prevent alternative protocols that exploit the many-particle nature of the working substance.

In the quantum domain, this state of affairs should be revisited as suggested by recent works focused on many-particle quantum thermodynamics.
The quantum statistics of the working substance can substantially affect the performance of a quantum engine \cite{Kim11}. 
The need to consider a many-particle thermodynamic cycle arises naturally in an effort to scale-up thermodynamic devices \cite{Rossnagel16,Maslennikov17,Serra18,Poschinger18} 
and has prompted the identification of  optimal confining potentials \cite{Zheng15} as well as the design of superadiabatic protocols \cite{Jaramillo16,Funo17,Deng18pra,Shujin18}, first proposed in a single-particle setting \cite{Deng13,delcampo14}.  The divergence of energy fluctuations in a working substance near a second-order phase transition has also been proposed to engineer critical Otto engines \cite{CF16}. 
In addition, many-particle thermodynamics can lead to quantum supremacy, whereby quantum effects boost the efficiency of a cycle beyond the classical achievable bound \cite{Jaramillo16,Reimann18}.   The realization of superadiabatic strokes with ultracold atoms using a Fermi gas as a working substance has been reported in \cite{Deng18pra,Shujin18,Diao18}.

Quantum technologies  have also uncovered novel avenues to design thermodynamic cycles. 
Traditionally, interactions between particles are generally considered to be ``fixed by Nature'' in condensed matter. 
However, a variety of techniques allow to modify interparticle interactions in different quantum platforms. 
A paradigmatic examples is the use of Feschbach and confinement-induced resonances in ultracold atoms \cite{Girardeau04}. Digital quantum simulation similarly allows to engineer interactions in  trapped ions and superconducting qubits \cite{Georgescu14}.

In this Letter, we introduce a novel thermodynamic cycle that exploits the many-particle nature of the working substance: it consists of four isochoric strokes,  alternating heating and cooling with a modulation of the interparticle interactions. This cycle resembles the four-stroke quantum Otto engine in that expansion and compression strokes are substituted by isochoric processes in which the interparticle interactions are modulated in time. Using Luttinger liquid theory, the efficiency of the cycle is shown to be universal in the low-temperature regime of a one-dimensional (1D) working substance. We further shown that when an interacting Bose gas is used as such, the average work output is maximum at criticality.

%
\begin{figure}[t]
\begin{center}
\includegraphics[width=0.8\linewidth]{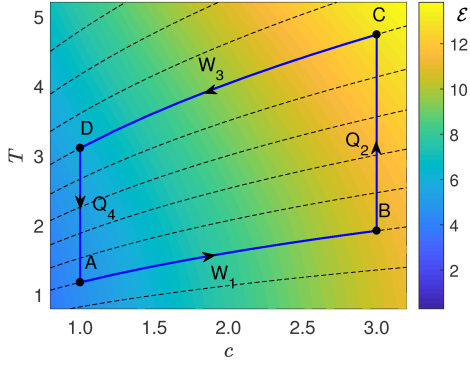}
\end{center}
\caption{\label{fig1}  {\bf Interaction driven quantum cycle.}  The working substance is driven through four sequential isochoric strokes alternating heating and cooling processes at different temperature $T$ with  isentropes in which the interparticle interaction strength $c$ is ramped up and down. The dashed lines correspond to the thermal entropy calculated from the thermodynamic Bethe ansatz equation  of the Lieb-Liniger gas  \cite{LL63,Lieb63,YY69,Yuzhu15}.} 
\end{figure}

{\it Interaction-driven thermodynamic cycle.---} We consider a quantum heat engine (QHE) with a working substance consisting of a low-dimensional ultracold gas tightly confined in a waveguide. Ultracold gases have been previously considered based  quantum  cycles  where work is done via expansion and compression processes, both in the non-interacting \cite{Saygin01,Zheng15} and interacting regimes \cite{Fialko12,Jaramillo16,Beau16,ChenBusch18,ChenDongSun18,Ma:2017}. 
We proposed the implementation of    quantum cycle consisting of four isochoric strokes, in which heating and cooling strokes are alternated with isentropic interaction-driven processes.  In the latter, work is done onto and by the working substance by increasing and decreasing the interatomic interaction strength, respectively. This work can be transferred to other degrees of freedom as we shall discuss below. The working substance consists of $N$ particles with interparticle interactions parameterized by the interaction strength $c$.
Both the particle number $N$ and system size $L$ are preserved throughout the cycle and any equilibrium point is parameterized by a point $(c,T)$ indexed by the temperature $T$ and interaction strength $c$.  
Specifically the interaction-driven quantum cycle,  shown in Fig.~\ref{fig1} for the 1D Lieb-Linger gas  \cite{LL63,Lieb63}, involves the following strokes:

(1) {\it Interaction ramp-up isentrope ($A\to B$):} The working substance is initially in the thermal state $A$ parameterized by $(c_A,T_A)$ and decoupled from any heat reservoir. Under unitary evolution, the interaction strength is enhanced to the value $c_B$ and the final state is non-thermal.
(2) {\it Hot isochore ($B\to C$):} Keeping $c_B$ constant, the working substance is put in contact with the hot reservoir at temperature $T_B$ and reaches the equilibrium state $(c_B,T_C)$.
(3) {\it Interaction ramp-down isentrope ($C\to D$):} The working substance in the equilibrium state $(c_B,T_C)$ is decoupled from the  hot reservoir and performs work
adiabatically while the interaction strength decreases from $c_B$ to $c_A$, reaching a non-thermal state. 
(4) {\it Cold isochore ($D\to A$):} The working substance is put in contact with the cold reservoir keeping the interaction strength constant until it reaches the thermal state $(c_A,T_A)$.
The work can be extracted from the heat engine is given by
$W=W_3-W_1=Q_2-Q_4$
while the efficiency of the heat engine reads
\begin{equation}\label{eq:efficience}
\eta=\frac{W}{Q_2}=1-\frac{Q_4}{Q_2}.
\end{equation}

{\it Interacting Bose gas as a working substance.---}
Consider  as a  working substance  an ultracold interacting Bose gas tightly confined in a waveguide \cite{Olshanii98,LSY03},  as realized in the laboratory \cite{Paredes05,Hofferberth08,Gring12}, see Fig.~\ref{fig1}. The effective Hamiltonian for $N$ particles is  that of the Lieb-Liniger  model \cite{LL63,Lieb63} 
\begin{equation}
\label{LLM}
\hat{H}_{\text{LL}}\!=\!-\sum_{j=1}^{N}\partial_{x_j}^{2}
+\sum_{1\le j<\ell\le N}2c\delta(x_{j\ell}),
\end{equation}
where $x_{j\ell}=x_j-x_{\ell}$, with $2m=\hbar=1$.
The spectral properties of Hamiltonian (\ref{LLM}) can be found using coordinate Bethe ansatz  \cite{LL63,Lieb63}. We consider a box-like trap \cite{Gaudin71,guan20051dbox} where any energy eigenvalue can be written as $E=\sum_ik_i^2$ in terms of the ordered the quasimomenta $0<k_1<k_2<\cdots<k_N$. The latter are  the (Bethe) roots $\{k_i\}$ of the coupled algebraic equations
\begin{equation}\label{eq:BA}
L k_i=\pi I_i-\sum_{j\neq i}\left(\arctan\frac{k_i-k_j}{c}+\arctan\frac{k_i+k_j}{c}\right),
\end{equation}
determined by the sequence of quantum numbers  $\{I_i\}$ with $i=1,\, 2,\, \cdots,\, N$.
As a function of $c$ and $T$, the 1D Bose gas exhibits a rich phase diagram.
We first consider the  strongly coupling regime and use a Taylor series expansion in $1/c$. For a given set of quantum numbers $\mathcal{I}_n=\{I_i^{(n)}\}$,  the corresponding energy eigenvalue takes the form
\begin{equation}
\label{EnergyStrong}
\epsilon_n(c)\approx \frac{\pi^2 \lambda_c}{L^2}\sum_{i=1}^N{I_i^{(n)} }^2,
\end{equation}
where the interaction-dependent factor $\lambda_c$ reads
\begin{equation}\label{lambda_c}
\lambda_c=1-\frac{4(N-1)}{c L}+\frac{12(N-1)^2}{c^2L^2}. 
\end{equation}
The spectrum of a strongly interacting Bose gas is thus characterized by eigenvalues with  scale-invariant behavior, i.e., 
$\epsilon_n(c)/\epsilon_n(c')=\lambda_c/\lambda_c'$. 
In this regime, the work output is thus set by $W = Q_2-Q_4 = [1-(\lambda_{c_A}/\lambda_{c_B})]Q_2$
and the efficiency
\beqa
\label{effstrong}
\eta=1-\frac{\lambda_{c_A}}{\lambda_{c_B}}
\eeqa
becomes independent of temperature of the heat reservoirs.

\begin{figure}[!t]
\centering
\includegraphics[width=0.8\linewidth]{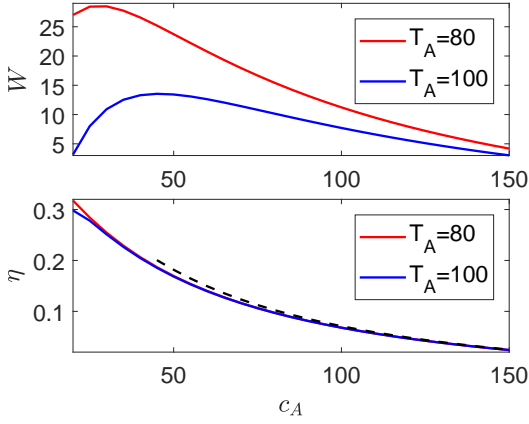}
\caption{{\bf Work and efficiency as functions of the interaction strength $c_A$.} The upper panel shows the dependence of the work output $W$ on $c_A$ for different $T_A$. In the lower panel, the solid lines show the efficiency $\eta$  obtained by numerical calculation, while the black dashed line shows the analytical result in Eq.~(\ref{effstrong}). Here, $N=5$, $L=1$, $T_C=150$, and $c_B=200$.}
\label{fig:Numerical}
\end{figure}

Beyond the strongly-interacting and low-energy regimes, we resort to a numerically-exact solution of Eqs.~(\ref{eq:BA}) for finite particle number $N$.  We  enumerate all the possible sets $\mathcal{I}_n$ of quantum numbers for low-energy states and solve Eqs.~(\ref{eq:BA}) numerically for a given interaction $c$. With the resulting quasi-momenta $\{k_{n,1},k_{n,2},\cdots\}$, and the corresponding energy eigenvalues  $\epsilon_n=\sum_i k_{n,i}^2$\,,  the probability that the Bose gas at temperature $T$ is found with energy $\epsilon_n$ is set by the Boltzmann weights $p_n=e^{-\epsilon_n/T}/\sum_m e^{-\epsilon_m/T}$ (with $k_B=1$). The equilibrium energy of the states $A$ and $C$ is set by thermal averages of the form $\la E\ra=\sum_np_n\epsilon_n$ that in turn yield the expressions for $Q_2$ and $Q_4$. Here, a proper cutoff of the possible sets $\{\mathcal{I}_n\}$ can be determined by $p_n/p_G \ll 1$, where $p_G$ is the probability for the working substance to be in the ground state. The numerical results for the efficiency $\eta$ and output work $W$ are shown in Fig.~\ref{fig:Numerical} as a function of the interaction strength.
For fixed values of $T_C$ and $T_A$,  the maximum  work  is studied as a function of $c_A$ while  keeping $c_B$ constant, see Fig.~\ref{fig:Numerical}.
The efficiency is well reproduced by  Eq. (\ref{effstrong}) at strong coupling, that captures as well a monotonic decay with increasing interaction strength. 
The efficiency is found to be essentially independent of the temperature in the strong interaction regime, whereas the work output is governed by  the temperature and interaction strength. 

 In the thermodynamic limit (where $N$ and $L \to \infty$ with $n=N/L$ being kept constant), the equilibrium state of the 1D Bose gas is determined by the Yang-Yang thermodynamics \cite{YY69,KBI93,Takahashi99,BatchelorGuan07,Yuzhu15}.  The pressure is then given by
\begin{equation}
p=\frac{T}{2\pi}\int \ln\left(1+e^{-\varepsilon(k)/T}\right)d k,
\label{pressureeq}
\end{equation}
where  the ``dressed energy''  $\varepsilon(k)$  is determined by thermodynamic Bethe Anstatz (TBA) equation
\begin{equation}
\varepsilon(k)=k^2-\mu-\frac{T}{2\pi}\int \frac{2 c}{c^2+(k-q)^2}\left(1+e^{-\varepsilon(q)/T}\right) d q. 
\label{TBA-E}
\end{equation}
The particle density $n$ and entropy density $s$ can be derived from the thermodynamics relations
\begin{equation}
n=\frac{\partial p}{\partial \mu},\quad s=\frac{\partial p}{\partial T},
\end{equation}
in terms of which the internal energy density reads  $\mathcal{E}=-p+\mu n+ T s$.  Both interaction-driven strokes are considered to be adiabatic.
As a result, the heat absorbed during the  hot isochore stroke ($B\to C$)
is given by $Q_2=L[\mathcal{E}(c_B,T_C)-\mathcal{E}(c_B,T_B)]$, while the heat  released during the cold isochore  ($D\to A$)
equals  $Q_4=L[\mathcal{E}(c_A,T_D)-\mathcal{E}(c_A,T_A)]$.
Here $T_B$ and $T_D$ can be determined from the entropy, by setting
$s(c_A,T_A)=s(c_B,T_B)$ and $s(c_B,T_C)=s(c_A,T_D)$, 
where $T_A$ and $T_C$ are the temperature of the cold and hot reservoir, respectively. The efficiency and work can then be obtained by numerically solving the TBA equation.

{\it Universal  efficiency at low temperature.---}
The low-energy behaviour  of 1D Bose gases is described  by the Tomonaga-Luttinger liquid (TLL) theory \cite{Luttinger63,MattisLieb65,Haldane80,Haldane81,Cazalilla04}, in which the free energy density reads
\begin{equation}
\mathcal{F}=\mathcal{E}_0-\frac{\pi T^2}{6 v_s},
\end{equation}
where $\mathcal{E}_0$ is the energy density of ground state, $v_s$ is the sound velocity which depends on particle density $n$ and interaction $c$. 
The entropy density $s$ can be obtained as the derivative of free energy, $s= -\partial\mathcal{F}/\partial T = \pi T / 3 v_s$. 
The expression for the heat absorbed and released are respectively given by
\begin{eqnarray}
Q_2&=&L \int_{s_B}^{s_C} T d s=\frac{\pi L}{6 v_s^B}(T_C^2-T_B^2)\,,\\
Q_4&=&L \int_{s_A}^{s_D} T d s=\frac{\pi L}{6 v_s^A}(T_D^2-T_A^2)\,,
\end{eqnarray}
where $s_i = \pi T_i/3v_s^i$ and $v_s^i$ with $v_s^B = v_s^C$ and $v_s^A = v_s^D$ are the entropy density and the sound velocity of the state $i$, respectively.
Using the fact that the strokes (1) and (3) are isentropes, it follows that 
\beqa
\xi\equiv\frac{T_A}{T_B}=\frac{T_D}{T_C}=\frac{v_s^A}{v_s^B}
\eeqa
 and as a result, the efficiency and work output are given by 
\begin{eqnarray}
\eta_{\rm TLL}  &=& 1-\frac{v_s^A}{v_s^B}=1-\xi, \label{effTLL} \\
W_{\rm TLL}&=& \frac{\pi L T_C^2}{6v_s^B}(1-\xi)\left(1-\frac{\kappa^2}{\xi^2}\right),
\label{eq:optimalW}
\end{eqnarray}
where $\kappa=T_A/T_C$. Since $T_A<T_B<T_C$, we have $0<\kappa<\xi<1$. 
The work output for fixed  $T_A$ and $T_C$ is maximized at $\xi=\xi_c$ as shown in \cite{SM},
\begin{equation}\label{xi_c}
\xi_c\simeq (2\kappa^2)^{\frac{1}{3}} [ 1-  \left(\kappa/2 \right)^{\frac{2}{3}}/3 ]\,.
\end{equation}

As the TLL theory describes the universal low-energy behavior of 1D many-body systems, 
Eqs.~(\ref{effTLL})--(\ref{xi_c}) provide a universal description of the efficiency and work of quantum heat engines with a 1D interacting working substance at low temperatures, which are applicable to any cycles whose working strokes and heat exchanging strokes are separated.
In particular, in the strongly interacting regime, the sound velocity of 1D Bose gases is given by $
v_s\simeq 2\pi n(1- 4n c^{-1} + 12n^2 c^{-2})$ \cite{Yuzhu15}. In this regime, thus the result (\ref{effTLL}) reduces to Eq.~(\ref{effstrong}).
On the other hand, in the weak interaction regime,
the sound velocity is given by
$v_s\simeq 2n\sqrt{\frac{c}{n}-\frac{1}{2\pi}\left(\frac{c}{n}\right)^{\frac{3}{2}}}$ \cite{Yuzhu15}, and thus the efficiency yields
\begin{equation}\label{eq:eta_s_weak}
\eta_{\rm sat}\simeq 1-\sqrt{\frac{c_A}{c_B}},
\end{equation}
indicating an enhancement of the performance with respect to the strongly-interacting case.

\begin{figure*}[t]
\centering
\includegraphics[scale=0.38]{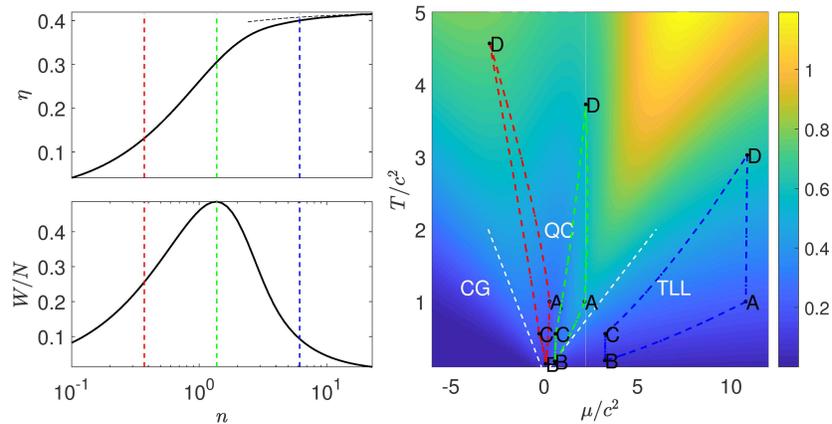}
\caption{{\bf Upper left panel:}  Efficiency $\eta$ as a function of the density $n$. {\bf Lower left panel:} average
work output $W /N$ as a function  of density $n$.  {\bf Right panel:} Phase diagram of Bose gases given by a contour plot of the  specific
heat. The three different polygons delineate interaction-driven thermodynamic cycles  in different regimes of the phase diagram corresponding to the three
dashed lines in the left panels. The equilibrium states A and C are chosen to be the same $c$ and $T$ for the different cycles.  The efficiency saturates at $0.42$ given by Eq.~(\ref{eq:eta_s_weak}) in the TLL regime.}
\label{fig:qhe_spc}
\end{figure*}

{\it Quantum critical region.---}
We next focus on the performance of an interaction-driven QHE at quantum criticality \cite{Yang17}. 
The 1D Bose gas displays a  rich critical behavior in the temperature $T$ and chemical potential $\mu$ plane, see Fig.~\ref{fig:qhe_spc}.  
In the region of $\mu\ll0$ and $T\gg n^2$, i.e. the mean distance between atoms is much larger  than thermal wave length, the system behaves as a classical gas (CG). A quantum critical region (QC) emerges between two critical temperatures (see the white dashed lines in Fig.~\ref{fig:qhe_spc}) fanning out from  the critical point $\mu_c=0$. In this region,  quantum  and thermal fluctuations have the same power of the temperature dependence.
In the region with $\mu>0$ and temperatures below the right critical temperature, the quantum and thermal fluctuations can reach equal footing: the TLL region discussed above.

For fixed cycle parameters ($c_A$, $c_B$, $T_A$, and $T_C$), we study the performance of the engine across the QC region by changing $n$.
We numerically calculate the efficiency $\eta$ and average work $W/N$  by using the TBA equation (\ref{TBA-E}), and show that near the quantum critical region $W/N$ has a maximum value, see Fig.~\ref{fig:qhe_spc}. We set $c_A=1$, $c_B=3$, $T_A=1$, and $T_C=5$ for the heat engine and let the density 
 $n$ increase from $0.1$ to $23$. The red, green, and blue dashed lines in Fig.~\ref{fig:qhe_spc} correspond to the density $n\simeq 0.2$, $1.4$, and $6.2$, respectively. In order to understand the maximum of $W/N$, we also plot this three engine cycles ($A\to B \to C \to D \to A$) in the phase diagram of specific heat in the $T-\mu$ plane in the right panel of Fig.~\ref{fig:qhe_spc}. When the engine works near the boundary from QC to TLL, it  has the maximum average work.

{\it Discussion.---} The modulation of the interaction strength $c$ is associated with the performance of work, that has recently been investigated for different working substances \cite{Chenu18,Chenu18b,Andrei18}.  
An experimentally-realizable work outcoupling mechanism can be engineered whenever the value of the coupling strength specifically depends on the configuration of other degrees of freedom.  This is analogous to the standard Carnot or Otto cycles where the working substance is confined in a box-like, harmonic, or more general potential \cite{Kosloff17}. The latter is endowed with a dynamical degree of freedom that is assumed to be slow (massive) so that in the spirit of the Born-Oppenheimer approximation can be replaced by a parameter.
Similarly, the modulation of the coupling strength in an interaction-driven cycle can be associated with the coupling to external degrees of freedom.

As an instance, consider the choice of an interacting SU(2) 1D spinor Fermi gas in a tight waveguide as a working substance. 
 The Hamiltonian can be mapped to the Lieb-Liniger model \cite{Girardeau06}
with an operator-valued coupling strength that depends on the spinor degrees of freedom $\hat{c}=\hat{c}(\hat{\mathbf{S}}_j\cdot\hat{\mathbf{S}}_{\ell})$.
The latter can be reduced to a real number $c=c(\la \hat{\mathbf{S}}_j\cdot\hat{\mathbf{S}}_{\ell}\ra)$ making use of a variational method 
 \cite{Girardeau06,delcampo07}, an approximation corroborated by the exact solution in a broad range of parameters \cite{Guan09}. Thus, the dependence of the interaction strength on the spin degrees of freedom provides a possible work outcoupling mechanism. 
An alternative  relies on the use of confinement-induced resonances \cite{Girardeau04,Haller10}. The scattering properties of a tightly confined quasi-1D working substance in a waveguide can be tuned by changing the transverse harmonic confinement of the waveguide. The frequency $\om_{\perp}$ of the latter directly determines the interaction strength $c$, i.e., $c=c(\om_\perp)$. In this case, the role of $\om_{\perp}$ parallels that of the box-size or harmonic frequency in the conventional Carnot and Otto cycles with a confined working substance.

Finally, we  note that an interaction-driven cycle can also be used to describe QHEs in which the interaction-driven strokes are substituted by processes involving the transmutation of the particle quantum exchange statistics, e.g., a change of the statistical parameter of the working substance. The Hamiltonian (\ref{LLM}) can be used to describe 1D anyons with pair-wise contact interactions with coupling strength $\tilde{c}$ and statistical parameter $\theta$  characterizing the exchange statistics, smoothly interpolating between bosons and fermions \cite{Kundu99,Batchelor06}. The spectral properties of this Lieb-Liniger anyons can be mapped to a bosonic Lieb-Liniger model (\ref{LLM}) with coupling strength  $c=\tilde{c}/\cos(\theta/2)$ \cite{Kundu99,Batchelor06}.  The modulation of $c$ can be achieved by the control of the particle statistics, tuning $\theta$ as proposed in \cite{Tassilo11}.

{\it Conclusions.---}
We have proposed an experimental realization of an  interaction-driven quantum heat engine that has no  single-particle counterpart: It is based on  a novel quantum cycle that alternates  heating and cooling strokes with processes that are both isochoric and isentropic and in which work is done onto or by the working substance by  changing in the interatomic interaction strength. This cycle can be realized  with a Bose gas in a tight-waveguide as a working substance. Using Luttinger liquid  theory, the engine efficiency has been shown to be   universal in the low temperature limit, and  set by the ratio of the sound velocities in the interaction-driven strokes. The optimal work can be achieved by changing the ratio of the sound velocity, e.g., by tuning the interaction strength. An analysis of the engine performance across the phase diagram of the Bose gas  indicates that quantum criticality maximizes the efficiency of the cycle.

Our proposal can be extended to Carnot-like interaction driven cycles in which work and heat are simultaneously exchanged in each stroke.
Exploiting effects beyond adiabatic limit  may lead to a quantum-enhanced performance \cite{Jaramillo16,Reimann18}.
The use of non-thermal reservoirs \cite{Scully03,Kieu04,Rossnagel14} and quantum measurements \cite{Watanabe17,Elouard17}  constitutes another interesting prospect.  Our results identify  confined Bose gases as an ideal platform for the engineering of scalable many-particle quantum thermodynamic devices.

{\it Acknowlegments.---} We thank F. J. G\'omez-Ruiz and Zhenyu Xu for comments on the manuscript.
Funding support from the John Templeton Foundation, the National Key R\&D Program of China No. 2017YFA0304500 and the key NSFC grant No. 11534014 and NSFC grant No. 11874393,  is  further acknowledged.
G.W. was supported by NSF of China (Grant No. 11674283), by the Fundamental Research Funds for the Central Universities (2017QNA3005, 2018QNA3004), by the Zhejiang University 100 Plan, and by the Thousand Young Talents Program of China.

\bibliography{IDQHE_biblio}

\newpage
\pagebreak

\clearpage
\widetext

\section{1D Bose gases in a hard wall potential}
Consider a system with $N$ identical bosons with a contact interaction confined in a one-dimensional (1D) hard-wall potential. This system is described by the  Hamiltonian
\begin{equation}
H=-\frac{\hbar^2}{2m}\sum_{i=1}^N \frac{\partial^2}{\partial x_i^2}+g_{\rm 1D}\sum_{1\leq i<j\leq N}\delta(x_i-x_j)\,,
\end{equation}
where $g_{\rm 1D}=\hbar^2 c/m$ is the 1D coupling constant and $c$ parametrizes the interaction strength. Hereafter, we set $\hbar=2m=1$. This model can be exactly solved by the Bethe ansatz \cite{Gaudin71,guan20051dbox}, and the wavefunction is given by
\begin{equation}
\Psi_{\{\epsilon_i k_i\}}(x_1,\cdots,x_N)=\sum_{\epsilon_1,\cdots,\epsilon_N}\sum_P \epsilon_1\cdots\epsilon_N\, A(\epsilon_1k_{P_1}, \cdots, \epsilon_N k_{P_N})\, e^{i(\epsilon_1 k_{P_1}x_1+\cdots+\epsilon_N k_{P_N}x_N)}\,,
\end{equation}
where the sum is taken over all $N!$ permutations $P$ of $N$ integers such that $P: (1, 2, \cdots, N) \rightarrow (P_1, P_2, \cdots P_N)$, and $\epsilon_i$ takes $\pm 1$. Here, $k_{P_i}$ is the the wave number which satisfies the Bethe equations
\begin{equation}
e^{i 2k_i L}=-\prod_{j=1}^N \frac{(k_i-k_j+i c)(k_i+k_j+i c)}{(k_i-k_j-i c)(k_i+k_j-i c)}\qquad (i=1, \cdots,\, N)\,,\label{eq:betheeq}
\end{equation}
and the energy of this state is given by 
\begin{equation}
E = \sum_{i=1}^N k_i^2\,.
\end{equation}
Taking the logarithm of the Bethe equations (\ref{eq:betheeq}), one  obtains
\begin{equation}\label{eq:BASM}
L k_i=\pi I_i-\sum_{j (\neq n)}\left(\arctan\frac{k_i-k_j}{c}+\arctan\frac{k_i+k_j}{c}\right)\qquad (i=1, \cdots,\, N)\,,
\end{equation}
where $\{I_i\}$ are integer quantum numbers, which are arranged in ascending order: $1\leq I_1\leq\cdots\leq I_N$. Thus, the wave numbers $\{k_i\}$ are also in ascending order, i.e., $0<k_1<\cdots<k_N$.

\subsection*{Scaling invariant behavior}
We next focus on the low energy excitations which dominate the thermodynamics in the strongly-interacting Bose gas at  low temperature.
A Taylor expansion of the rhs of Eq.~(\ref{eq:BASM}) for $c\gg k_N$ yields the following asymptotic solution 
\begin{eqnarray}
L k_i & = & \pi I_i -\sum_{i (\neq n)}^N\left[\frac{k_i-k_j}{c}+\frac{k_i+k_j}{c}\right]+O\left(\frac{k_N^3}{c^3}\right)\nonumber \\
	  & = & \pi I_i +\frac{2(1-N)k_i}{c}+O\left(\frac{k_N^3}{c^3}\right)\nonumber\\
	  & \approx & \pi I_i +\frac{2(1-N)}{c}\left[\frac{\pi I_i}{L}+\frac{2(1-N)\pi I_i}{cL^2}\right]+O\left(\frac{k_N^3}{c^3}\right)\nonumber\\
	  & \approx & \pi\left[1+\frac{2(1-N)}{c L}+\frac{4(1-N)^2}{c^2L^2}\right]I_i+O\left(\frac{k_N^3}{c^3}\right). 
\end{eqnarray}
Note that, in this regime, the dependence of the quasimomentum $k_i$ on the other quasimomenta $k_j$ with $j\neq i$ is of higher order in $k_N/c$. In this sense, the algebraic Bethe ansatz equations (\ref{eq:BA}) decouple in this limit.

The energy eigenvalue $\varepsilon_n(c)$ for a given set of quantum numbers $\mathcal{I}_n \equiv \{I_i^{(n)}\}$ is set by
\begin{eqnarray}
\varepsilon_n(c) & \approx &\sum_{i=1}^N k_i^2\nonumber \\
 			& \approx & \frac{\pi^2}{L^2}\left[1-\frac{4(N-1)}{c L}+\frac{12(N-1)^2}{c^2L^2}\right]\sum_{i=1}^N{I_i^{(n)}}^2+\frac{1}{L^2} O\left(\frac{k_N^3}{c^3}\right).\label{eq:eigenenergy}
\end{eqnarray}
Thus, up to corrections of order  $1/c^2$,  the energy eigenvalue can be written in terms of that of free fermions rescaled by  an overall factor resulting from the interactions. We refer to this factor as the generalized exclusion statistics parameter, in view of \cite{BatchelorGuan07},  and define it as
\begin{equation}
\lambda_c=1-\frac{4(N-1)}{c L}+\frac{12(N-1)^2}{c^2L^2}\,.
\end{equation}
Using it,  the energy eigenvalues of a strongly-interacting 1D Bose gases simply read
\begin{equation}
\label{EnergyStrong}
\varepsilon_n(c)\approx \frac{\pi^2 \lambda_c}{L^2}\sum_{i=1}^N {I_i^{(n)}}^2\,,
\end{equation}
and exhibit the  scale-invariant behavior
\begin{equation}
\label{eq:EnergyStrong2}
\frac{\varepsilon_n(c)}{\varepsilon_n(c')}=\frac{\lambda_c}{\lambda_{c'}}. 
\end{equation}

\section{Energies of equilibrium and non-equilibrium states}
The equilibrium state of the system at temperature $T$ can be characterized   by the  partition function
\begin{equation}
Z=\sum_{n=1}^{\infty}e^{-\varepsilon_n(c)/T}\,,
\end{equation}
where the summation is taken over all the quantum states. The average energy of the system is given by
\begin{equation}
\varepsilon_{\rm eq}(c,T)=\sum_{n=1}^{\infty} p_n(c,T)\, \varepsilon_n(c)
\end{equation}
with $p_n(c,T)$ being the probability measure of the $n$-th eigenstate in the Gibbs ensemble given by
\begin{equation}\label{eq:probality}
p_n(c,T)=e^{-\varepsilon_n(c)/T}/Z\,.
\end{equation}

During the interaction-driven strokes, the system generally deviates from the initial equilibrium state as a result of the modulation of the interaction strength $c$. However, since the energy gap for low-energy excitations is nonzero in the system with finite $N$ bosons, a sufficiently slow ramping process  becomes adiabatic and keeps the probability distribution $p_n(c,T)$ unchanged. Therefore, for the non-equilibrium state resulting from  an adiabatic ramping process, the average energy of the system at the interaction parameter $c'$ is given by
\begin{equation}
\varepsilon_{\rm neq}(c';c,T)=\sum_{n=1}^{\infty} p_n(c,T)\, \varepsilon_n(c').
\end{equation}
Due to the scale-invariant behavior at strong coupling,  the energy of the non-equilibrium state at the end of the interaction-ramp isentropic process is found to be
\begin{equation}\label{eq:eq2neq}
\varepsilon_{\rm neq}(c';c,T)=\frac{\lambda_{c'}}{\lambda_c}\, \varepsilon_{\rm eq}(c,T).
\end{equation}

Furthermore, the scale-invariant behavior Eq.~(\ref{eq:EnergyStrong2}) indicates that we can relate a non-equilibrium state in  adiabatic interaction-driven processes to an equilibrium state with an effective temperature $T'$ by noting that
\begin{eqnarray}
\varepsilon_{\rm neq}(c';c,T) & = & \sum_{n=1}^{\infty} \frac{e^{-\varepsilon_n(c)/T}}{\displaystyle \sum_{m=1}^{\infty}e^{-\varepsilon_m(c)/T}}\, \varepsilon_n(c')\nonumber\\
& = & \sum_{n=1}^{\infty}\frac{e^{-\varepsilon_n(c')\lambda_c/T\lambda_{c'}} }{\displaystyle \sum_{m=1}^{\infty}e^{-\varepsilon_m(c')\lambda_c/T\lambda_{c'}}}\, \varepsilon_n(c')\nonumber\\
& = & \varepsilon_{\rm eq}(c',T')\,,
\end{eqnarray}
where the effective temperature  reads 
\begin{equation}
T'=\lambda_{c'}T/\lambda_c.
 \end{equation}

\section{Quantum heat engine}
For the interaction-driven quantum heat engine (see Fig. 1 in the main text), the heat absorbed from the hot reservoir is given by
\begin{eqnarray}
Q_2 & = & \varepsilon_{\rm eq}(c_B,T_C)-\varepsilon_{\rm neq}(c_B;c_A,T_A)\nonumber\\
   & = & \varepsilon_{\rm eq}(c_B,T_C)-\frac{\lambda_{c_B}}{\lambda_{c_A}}\, \varepsilon_{\rm eq}(c_A,T_A)\,,\label{eq:q1}
\end{eqnarray}
while the heat released from the engine to the cold reservoir reads
\begin{eqnarray}
Q_4 & = & \varepsilon_{\rm neq}(c_A;c_B,T_C)-\varepsilon_{\rm eq}(c_A,T_A)\nonumber\\
   & = & \frac{\lambda_{c_A}}{\lambda_{c_B}}\, \varepsilon_{\rm eq}(c_B,T_C)-\varepsilon_{\rm eq}(c_A,T_A)\,.\label{eq:q2}
\end{eqnarray}
Therefore, the efficiency $\eta$ and the work $W$ done by the engine are given by
\begin{eqnarray}
  \eta &=& 1-\frac{Q_4}{Q_2}=1-\frac{\lambda_{c_A}}{\lambda_{c_B}}\,,\label{efficiency}\\
  W &=& Q_2-Q_4 = \left(1-\frac{\lambda_{c_A}}{\lambda_{c_B}}\right)Q_1\,.\label{work-a}
\end{eqnarray}

The efficiency $\eta$ and work $W$ can also be obtained numerically by solving Eq.~(\ref{eq:BA}). In our calculations, we take a proper cutoff for excited states whose probability $p_n$ given by Eq.~(\ref{eq:probality}) is much smaller than that of the ground state.

\section{Thermodynamic limit}

In the laboratory,  a Bose gas confined in an effectively 1D trap typically consists of thousands of particles. It is generally considered that  such system is well described by the thermodynamic limit, i.e., $N$ and $L \to \infty$ with $n=N/L$ being kept constant.
The equilibrium states can then be described by the Yang-Yang thermodynamic equation \cite{YY69}. The pressure $p$, particle density $n$, entropy density $s$  and internal energy density $\mathcal{E}$ can then be found as described in the text.


Note that  we take the thermodynamic limit after the slow ramping limit to guarantee that no diabatic transitions occur during the ramping processes. Thus, the probability $p_n(c,T)$ is unchanged during the ramping processes.
In these processes, the entropy should also be constant because no heat is transferred in or out of the working substance. Given the expressions for the heat absorbed during the hot isochore stroke ($B$ to $C$) 
\begin{equation}
Q_2=L\left[\mathcal{E}(c_B,T_C)-\mathcal{E}(c_B,T_B)\right]\,
\end{equation}
and the heat released during the cold isochore stroke ($D$ to $A$)
\begin{equation}
Q_4=L\left[\mathcal{E}(c_A,T_D)-\mathcal{E}(c_A,T_A)\right]\,,
\end{equation}
the efficiency $\eta$ and work output $W$ are given by
\begin{eqnarray}
  \eta &=& 1-\frac{\mathcal{E}(c_A,T_D)-\mathcal{E}(c_A,T_A)}{\mathcal{E}(c_B,T_C)-\mathcal{E}(c_B,T_B)}\,,\\
  W &=& L\left[\mathcal{E}(c_B,T_C)-\mathcal{E}(c_B,T_B)- \mathcal{E}(c_A,T_D)+\mathcal{E}(c_A,T_A)\right]\,.
\end{eqnarray}
Here, $T_B$ and $T_D$ can be determined using the fact that the interaction-driven strokes are isentropic, namely
\begin{equation}\label{eq:isentropy}
s(c_A,T_A)=s(c_B,T_B),\quad s(c_B,T_C)=s(c_A,T_D)\,,
\end{equation}
where $T_A$ and $T_C$ denote the temperature of the cold and hot reservoir, respectively. The efficiency can thus be obtained numerically by solving Eq.~(\ref{TBA-E}).

\subsection*{Universality at low energies}
The low-energy physics of 1D Bose gases can be well captured by the Luttinger liquid theory. The free energy density $\mathcal{F}$ is given by \cite{guan2011polylogs}
\begin{equation}
\mathcal{F}\approx\mathcal{E}_0-\frac{\pi T^2}{6 v_s}\,,
\end{equation}
where $\mathcal{E}_0$ is the energy density of the ground state, $v_s$ is the sound velocity which depends on the particle density $n$ and the interaction strength $c$. The entropy density $s$ is given by the derivative of the free energy, namely,
\begin{equation}
s=-\frac{\partial\mathcal{F}}{\partial T}=\frac{\pi T}{3 v_s}. 
\end{equation}
The heat absorbed from the hot resevoir in the hot isochore stroke ($B$ to $C$) is given by
\begin{equation}
Q_2=L \int_{s_B}^{s_C} T d s=\frac{\pi L}{6 v_s^B}(T_C^2-T_B^2)\,,
\end{equation}
and the heat released to the cold reservoir in the cold isochore stroke ($D$ to $A$) is
\begin{equation}
Q_4=L \int_{s_A}^{s_D} T d s=\frac{\pi L}{6 v_s^A}(T_D^2-T_A^2).
\end{equation}
Here, $s_x = \pi T_x/3v_s^x$ and $v_s^x$ with $v_s^B = v_s^C$ and $v_s^A = v_s^D$ are the entropy density and the sound velocity of the state $x \in \{A, B, C, D\}$, respectively.

\begin{figure}[!t]
\centering
\includegraphics[scale=0.5]{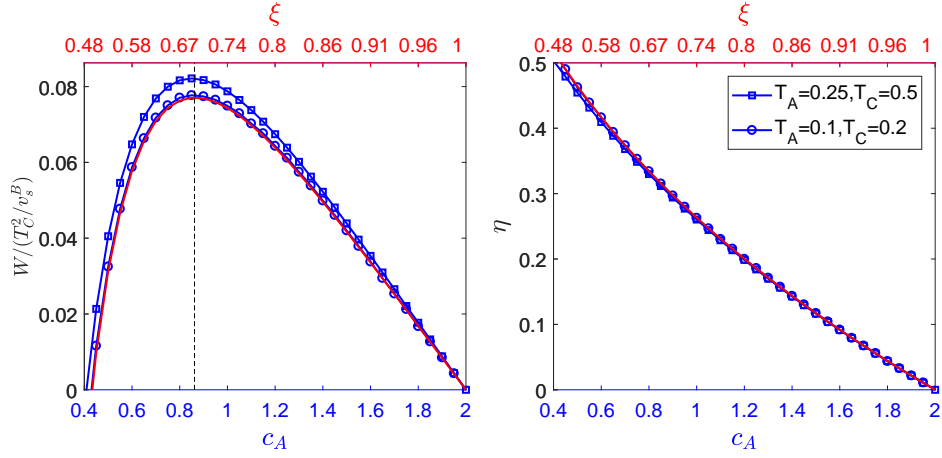}
\caption{{\bf Work $W$ and efficiency $\eta$ vs interaction strength $c_A$.} The blue lines with squares and circles are obtained by numerically solving the TBA equation. The red solid lines in the left and right panels are obtained by Eqs.~(\ref{eq:optimalW}) and (\ref{eq:tllefficiency}), respectively, where $\xi$ for a given $c_A$ is obtained by numerically calculating the sound velocities $v_s^A$ and $v_s^B$ at zero temperature \cite{Yuzhu15}. The black dashed line corresponds to $\xi_c\approx 0.69$ for $\kappa=0.5$ given by Eq.~(\ref{eq:xi_c}). In our numerical calculation, we set $n=1$, $c_B=2$, and $L=1$.}  
\label{fig:QHE_tll}
\end{figure}

By Eq.~(\ref{eq:isentropy}), we obtain
\begin{equation}
\frac{\pi T_A}{3 v_s^A}=\frac{\pi T_B}{3 v_s^B},\quad \frac{\pi T_D}{3 v_s^A}=\frac{\pi T_C}{3 v_s^B}\,,
\end{equation}
namely,
\begin{equation}
\frac{T_A}{T_B}=\frac{T_D}{T_C}=\frac{v_s^A}{v_s^B}\,.
\end{equation}
To simplify, we introduce the following two dimensionless parameters:
\begin{equation}
\xi=\frac{v_s^A}{v_s^B},\quad \kappa=\frac{T_A}{T_C}\,.
\end{equation}
Since $T_A<T_B<T_C$, we get
\begin{equation}
0<\kappa<\xi<1
\end{equation}
The work can be extracted from the heat engine is given by 
\begin{eqnarray}
W & = & Q_2-Q_4\nonumber\\
  & = & \frac{\pi L}{6v_s^B}(T_C^2-T_B^2)-\frac{\pi L}{6v_s^A}(T_D^2-T_A^2)\nonumber\\
  & = & \frac{\pi L T_C^2}{6v_s^B}(1-\xi)\left(1-\frac{\kappa^2}{\xi^2}\right)\,.
  \label{eq:optimalW}
\end{eqnarray}
An interesting question is what is the optimal work extracted from this heat engine when the temperatures of the two reservoirs, $T_A$ and $T_C$, are fixed.
It is easy to show that the work output $W$ always has a maximum value (see the right panel of Fig.~\ref{fig:QHE_tll}) at $\xi = \xi_c$ with
\begin{equation}
\xi_c=\frac{\kappa^2}{a^{1/3}}-\frac{a^{1/3}}{3}\,,
\end{equation}
where $a=27 \kappa^2 \left[\sqrt{1+(\kappa^2/27)}-1\right] \approx \kappa^4/2$.
Thus for small $\kappa \ll 1$, we get
\begin{equation}\label{eq:xi_c}
\xi_c\approx (2\kappa^2)^{\frac{1}{3}}\left[1-\frac{1}{3}\left(\frac{\kappa}{2}\right)^{\frac{2}{3}}\right]\,.
\end{equation}

The efficiency of the heat engine is 
\begin{eqnarray}\label{eq:tllefficiency}
\eta & = & 1-\frac{Q_4}{Q_2}\nonumber\\
     & = & 1-\frac{v_s^B}{v_s^A}\frac{T_D^2-T_A^2}{T_C^2-T_B^2}\nonumber\\
     & = & 1-\xi\,.
\end{eqnarray}
Especially, for the strong interaction case, the sound velocity is $v_s\approx 2\pi n \left[1- 4 (n/ c) + 12 (n/c)^2 \right]$ \cite{Yuzhu15}, and thus the efficiency is given by
\begin{equation}
\eta \approx 1-\frac{1-\frac{4n}{c_A}+\frac{12 n^2}{c_A^2}}{1-\frac{4n}{c_B}+\frac{12 n^2}{c_B^2}}\,,
\end{equation}
which agrees with Eq.~(\ref{efficiency}) in the thermodynamic limit. For the weak interaction case, the sound velocity is $v_s\approx 2n \left[ (c/n) - (2\pi)^{-1} (c/n)^{3/2} \right]^{1/2}$ \cite{Yuzhu15}, and the efficiency is given by
\begin{equation}
\eta\approx 1-\sqrt{\frac{c_A}{c_B}}\,.
\end{equation}

The work output and the efficiency given by Eqs.~(\ref{eq:optimalW}) and (\ref{eq:tllefficiency}) are universal at low energies for 1D system.  Here we shall take 1D Bose gases as a platform to test these universal properties. For the interaction-driven engine with fixed particle number $N$ and the length $L$ studied in the present work, the sound velocity is changed during the two isoentropic strokes [($A$ to $B$) and ($C$ to $D$)] by changing the interaction strength. We numerically calculate the work $W$ and efficiency $\eta$ for $\kappa=0.5$ and $n=N/L=1$. The interaction strength $c_B$ is fixed at $c_B=2$, which corresponds to the sound velocity $v_s^B\approx 2.50$. From Eq.~(\ref{eq:xi_c}), the optimal work is obtained at
\begin{equation}
\xi_c \approx 0.69
\end{equation}
for $\kappa=0.5$.
The numerical results are shown by the blue lines in Fig.~\ref{fig:QHE_tll}. In the low temperature region, the numerical results are well explained by the results of the Luttinger liquid theory given by Eqs.~(\ref{eq:optimalW}) and (\ref{eq:tllefficiency}), which are shown by the red solid lines. In the high temperature region, these two results deviate, which indicates the breakdown of the Luttinger liquid theory.

\subsection*{Quantum criticality}

\begin{figure}[!t]
\centering
\includegraphics[scale=0.35]{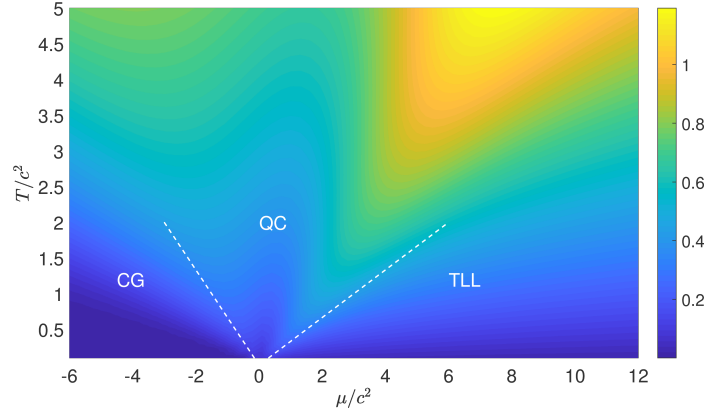}
\caption{{\bf Phase diagram of the 1D Bose gas.} The color contour shows the specific heat. With increasing the chemical potential $\mu$, the system undergoes a crossover from the classical gas (CG) to the Tomonaga-Luttinger liquid (TLL) across the quantum critical (QC) region.
}
\label{fig:spc}
\end{figure}

The 1D Bose gases show rich critical properties in the plane of the temperature $T$ and the chemical potential $\mu$; see Fig.~\ref{fig:spc}. In the so-called the Tomonaga-Luttinger liquid (TLL) region, where $\mu>0$ and the temperature is below the right critical temperature, the magnitude of quantum and thermal fluctuations are comparable. In this region, the low-energy properties can be well captured by the Luttinger liquid theory and the excitation spectrum is given by $E=v_s |k|$, where $v_s$ and $k$ are the sound velocity and the wave vector, respectively.
In the region of $\mu/T\ll 0$, i.e., the mean distance between atoms is much larger than thermal wave length, the system behaves as a classical gas (CG). A quantum critical (QC) region emerges between two critical temperatures (see the white dashed lines in Fig.~\ref{fig:spc}) fanning out from  the critical point $\mu_c=0$. In this region, quantum fluctuation and thermal fluctuation have the same power of the temperature dependence.  The density $n$ is monotonically increasing with the chemical potential $\mu$ for a given temperature $T$ and the  interaction strength $c$. Therefore, one can go across the QC region from the CG to TLL regions only by increasing the density $n$ of the working substance. Our numerical calculation shows $W/N$ takes a maximum value in the crossover region between QC and TLL (see Fig.~3 in the main text).

\section{1D Spinor Fermi gas as working substance}
Denoting by 
$\hat{P}_{ij}^s=\frac{1}{4}-\hat{\mathbf{S}}_i\cdot\hat{\mathbf{S}}_{j}$ 
and 
$\hat{P}_{ij}^t=\frac{3}{4}+\hat{\mathbf{S}}_i\cdot\hat{\mathbf{S}}_{j}$
are the projectors onto the subspaces of singlet and triplet functions of the 
spin arguments $(\sigma_i,\sigma_j)$ for fixed values of all other 
arguments,  the Hamiltonian of the system is given by \cite{Girardeau06,delcampo07}
\begin{equation}\label{SFGH}
\hat{H}_{\text{SFG}}=-\sum_{i=1}^{N}\partial_{x_i}^{2}
+\sum_{1\le i<j\le N}[g_{e}\delta(x_{ij})\hat{P}_{ij}^s
+v_{o}(x_{ij})\hat{P}_{ij}^t]\ .
\end{equation}
Here,  $v_{o}$ is a strong, attractive, zero-range, and
odd-wave interaction  that is the 1D analog of 3D p-wave interaction.
Similarly, $g_{e}$ denotes  the even-wave 1D coupling constant   arising from 3D s-wave scattering
\cite{Olshanii98}.
The Hamiltonian of the spinor Fermi gas can be mapped to the Lieb-Liniger Hamiltonian \cite{Girardeau06}
by promoting the coupling strength to an operator dependent on the spin degrees of freedom
\begin{equation}
\hat{c}= 2^{-1} \left[ (3c_o+c_e)/2
+2(c_o-c_e)\hat{\mathbf{S}}_i\cdot\hat{\mathbf{S}}_{j}\right],
\end{equation} 
where the coupling constants $\{c_o,c_e\}$ are set by 
$g_{e}$ and $v_{o}$.
The operator-valued coupling strength $\hat{c}$ can however be reduced to a real number $c=(3c_o+c_e)/4
+(c_o-c_e)\la \hat{\mathbf{S}}_i\cdot\hat{\mathbf{S}}_{j}\ra$ making use of a variational method that combines
the exact solution of the LL model with that of the 1D Heisenberg models  \cite{Girardeau06,delcampo07}. As it turns out, this approximation is corroborated by the exact solution of (\ref{SFGH}) in a broad range of parameters \cite{Guan09}. Thus, the dependence of the interaction strength on the spin degrees of freedom provides a possible work outcoupling mechanism.

\end{document}